\documentclass{ws-p9-75x6-50}
\usepackage{latexsym}
\usepackage{amsfonts}
\def\be{\begin{equation}}
\def\ee{\end{equation}}

\begin{document}

\title{Algebraic integrability of FRW-scalar cosmologies}

\author{Spiros Cotsakis}

\address{GEODYSYC, Department of Mathematics, University of the
Aegean,  83200, Samos, Greece\\E-mail: skot@aegean.gr}

\author{John Miritzis}

\address{Department of Marine Sciences, University of the Aegean, Mytilene 81100, Greece\\
E-mail: john@env.aegean.gr}


\maketitle

\abstracts{For dynamical systems of dimension three or more the
question of integrability or nonintegrability is extended by the
possibility of chaotic behaviour in the general solution. We
determine the integrability of isotropic cosmological models in
general relativity and string theory with a variety of matter
terms, by a performance of the  Painlev\'{e} analysis in an effort
to examine whether or not there exists a Laurent expansion of the
solution about a movable pole which contains the number of
arbitrary constants necessary for a general solution.}

The question of integrability in the context of cosmological
dynamical systems is intimately linked with their asymptotic
states for small or large times and  is obviously connected with
the nature of the singularities present in homogeneous
cosmologies. This question cannot be fully addressed via special
exact solutions since they are not easily matched to characterize
generic features of the models. Qualitative methods, on the other
hand, may serve such a purpose and the so-called Painlev\'{e}
analysis is one such approach for testing integrability in an
algebraic sense.

In this Research Announcement, a start is made of the Painlev\'{e}
analysis of different cosmologies that arise in general relativity
and string theory, in an effort to shed light to the question of
whether or not integrability is a generic property characterizing
cosmological dynamics (for the first results in this connection
see\cite{co-le-mi}).

1. The two-fluid FRW model in general relativity is described by
the system,
\begin{eqnarray}
\dot{\Omega}  &=&
-\mbox{$\frac{1}{2}$}(b-x)\cos2\Omega\cos\Omega\nonumber\\
\dot{\chi}  &=&(1-\chi^{2})\sin\Omega, \label{1.7}%
\end{eqnarray}
where the so-called compactified density parameter,
$\Omega\in\lbrack
-\mbox{$\frac{1}{2}$}\pi,\mbox{$\frac{1}{2}$}\pi],$ the transition
variable $\chi\in\lbrack-1,1]$ describes which fluid is dominant
dynamically and $b>-1$. The system (\ref{1.7}) is not in a
suitable form for the application of the Painlev\'{e} analysis,
but in the  new variables $x=\chi , y=\sin\Omega$, we can tranform
the system to the form,
\begin{eqnarray}
\dot{x}  &=&(1-x^{2})y\nonumber\\ \dot{y}
&=&-\mbox{$\frac{1}{2}$}(b-x)(1-2y^{2})(1-y^{2}).
\end{eqnarray}
In the expansions we have obtained, only one has the required
number of arbitrary constants and we cannot conclude that the
system (\ref{1.7}) is integrable in the sense of Painlev\'{e}.
However, it is possible to obtain a first integral of the original
system (\ref{1.7}) and reduce the solution to a rather complicated
quadrature. In fact, the system  is Hamiltonian so that the
existence of the first integral immediately guarantees
integrability in the sense of Liouville.

2. The flat one-fluid FRW space-time with a scalar field with an
exponential potential in general relativity\cite{gr-qc/9711068}
which reduces to a two-dimensional system is integrable in the
sense of Painlev\'{e}.

3. General relativistic one-fluid FRW models with $n$ scalar fields $\phi_{i}%
$, $i=1,\dots ,n$ with exponential potentials\cite{gr-qc/9911075}
expressed as three-dimensional systems prove to be integrable.

4. The four dimensional flat string FRW model with negative
central charge deficit described in the compactifying variables
$(\xi,\eta)$ by a two dimensional system\cite{gr-qc/9903095}
passes the ``weak'' Painlev\'{e} test and so is integrable in the
sense of Painlev\'{e}.

5. The ten-dimensional flat string FRW model in the RR sector and
with positive cosmological constant described by a
three-dimensional system\cite{gr-qc/9910074} passes the
Painlev\'{e} test and so is integrable in the sense of
Painlev\'{e}.

Among the models discussed in this Note, the two-fluid FRW model
in general relativity presents the most interesting dynamical
behaviour since it possesses the interesting property of the
so-called \emph{singular envelopes}, first discussed by Ince and
rediscovered in a different context in\cite{Cotsakis}. That is,
although the general solution is unknown, any possible
nonintegrable or even chaotic behaviour may be confined to that
region of phase space enveloped by the peculiar solutions. All
other models are either strictly integrable in the sense of
possessing the strong Painlev\'{e} property or have branch point
singularities indicating weak integrability in the sense of
Painlev\'{e}.

An interesting question is whether integrability in the case of
scalar field models is maintained when one considers more general
potentials. This is known not to be the case even for a general
FRW model with a scalar field that has a simple quadratic
potential. One would like to be able to relate the integrability
properties of different cosmological spacetimes in the context of
different gravity theories and matter fields in an effort to
understand the significance of exceptional non-integrable cases as
opposed to generic, integrable ones in the simple frame of
isotropic models before one moves on to the more difficult
homogeneous but anisotropic case. Problems in this direction are
currently under investigation and results will be given elsewhere.


\begin{thebibliography}{9}
\bibitem{co-le-mi}J. Miritzis \textit{et al, Symmetry, Singularities and
Integrability in Complex Dynamics IV: Painleve Integrability of
Isotropic Cosmologies}, (preprint: gr-qc/0011019)

\bibitem{gr-qc/9711068}E.J. Copeland, A.R. Liddle  and D. Wands \textit{Phys.Rev.D.}
\textbf{57}, 4686
 (1998)

\bibitem{gr-qc/9911075}A.A. Coley and R.J. van der Hoogen, \textit{The dynamics of
multi-scalar field cosmological models and assisted inflation},
(preprint: gr-qc/9911075)

\bibitem{gr-qc/9903095}A.P. Billyard, A.A. Coley and J.E. Lidsey, \textit{Qualitative
analysis of string cosmologies}, (preprint: gr-qc/9903095)

\bibitem{gr-qc/9910074}A.A. Coley, \textit{Dynamical systems in cosmology},
 (preprint: gr-qc/9910074)

\bibitem{Cotsakis}S. Cotsakis and P.G.L. Leach,
\textit{J Phys A: Math Gen} \textbf{27}, 1625 (1994)









\bibitem{wa-el}J. Wainwright and G.F.R. Ellis,  \textit{Dynamical Systems in
Cosmology} (Cambridge University Press, Cambridge, 1997)
\end{thebibliography}
\end{document}